\documentclass[a4paper]{article}
\usepackage{Odyssey2020}
\usepackage{epsfig,amssymb,amsmath}
\usepackage{multirow}
\usepackage{multicol}
\usepackage{color}
\usepackage{url}
\usepackage{subfig} 
\ninept

\title{Subband modeling for spoofing detection in automatic speaker verification}

\name{Bhusan Chettri$^{1,2}$, Tomi Kinnunen$^1$, Emmanouil Benetos$^2$ \thanks{This work was supported in part by the Academy of Finland (Proj. No. 309629 --- entitled “NOTCH: NOn-cooperaTive speaker CHaracterization”). EB is supported by RAEng Research Fellowship RF/128 and a Turing Fellowship.}}

\address{$^1$School of Computing, University of Eastern Finland, FI-80101, Joensuu, Finland \\ $^2$School of EECS, Queen Mary University of London, United Kingdom}

\begin{document}
\maketitle

\begin{abstract}
Spectrograms --- time-frequency representations of audio signals --- have found widespread use in neural network-based spoofing detection. While deep models are trained on the fullband spectrum of the signal, we argue that not all frequency bands are useful for these tasks. In this paper, we systematically investigate the impact of different subbands and their importance on replay spoofing detection on two benchmark datasets: ASVspoof 2017 v2.0 and ASVspoof 2019 PA. We propose a joint subband modelling framework that employs $n$ different sub-networks to learn subband specific features. These are later combined and passed to a classifier and the whole network weights are updated during training. Our findings on the ASVspoof 2017 dataset suggest that the most discriminative information appears to be in the first and the last $1$ kHz frequency bands, and the joint model trained on these two subbands shows the best performance outperforming the baselines by a large margin. However, these findings do not generalise on the ASVspoof 2019 PA dataset. This suggests that the datasets available for training these models do not reflect real world replay conditions suggesting a need for careful design of datasets for training replay spoofing countermeasures.

\end{abstract}


%
\section{Introduction}

\emph{Automatic speaker verification} (ASV) \cite{reynolds_SC1995} systems, similar to other biometric modalities, are prone to being intentionally fooled using \emph{spoofing attacks} \cite{Wu2015_survey} (or \emph{presentation attacks} \cite{isopad}), such as replay, text-to-speech (TTS), and voice conversion (VC). High-stakes ASV applications demand trustworthy fail-safe mechanisms (countermeasures) against such attacks. Here, a \emph{countermeasure} (CM) is defined as a binary classifier that aims at discriminating \emph{bonafide} (human speech) utterances from spoofing attacks. To allow maximum re-usability across different applications, the ideal CM should generalize across environments, speakers, languages, channels, and attacks. In practice, this is not the case; CMs are prone to overfitting. This could be due to variations within the spoof class (\emph{e.g.} speech synthesizers not present in the training set), within the bonafide class (\emph{e.g.} due to content and speaker), or extrinsic nuisance factors (\emph{e.g.} background noise). 

\begin{figure}
	\subfloat[Baseline. Traditional CNN model trained on the fullband spectrum. ]{\includegraphics[width=\linewidth]{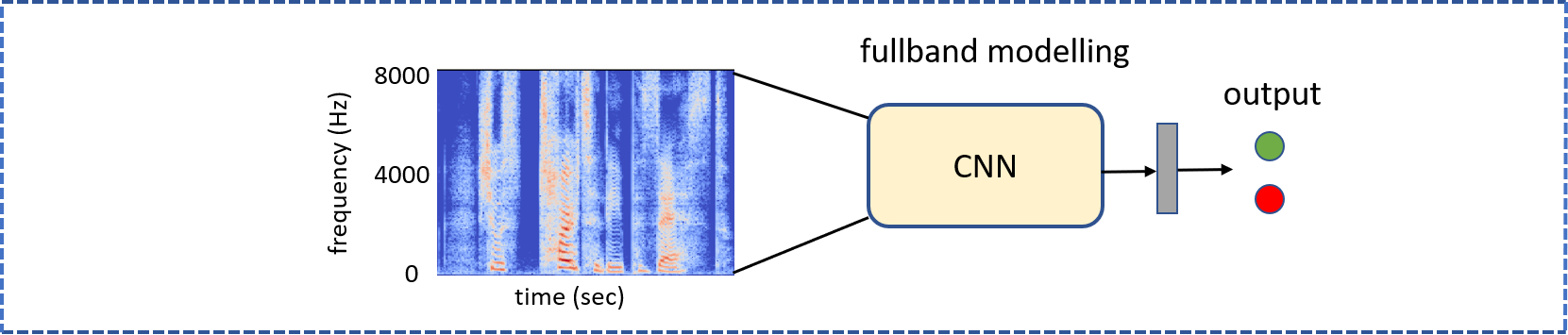}} \\
	\subfloat[Proposed framework. (i) The original spectrogram is split into $n$ sub-spectrograms on which $n$ independent CNNs are trained. (ii) uses pretrained weights from (i) to initialise $n$ subband CNNs and the whole network weights are updated during training.]{\includegraphics[width=\linewidth]{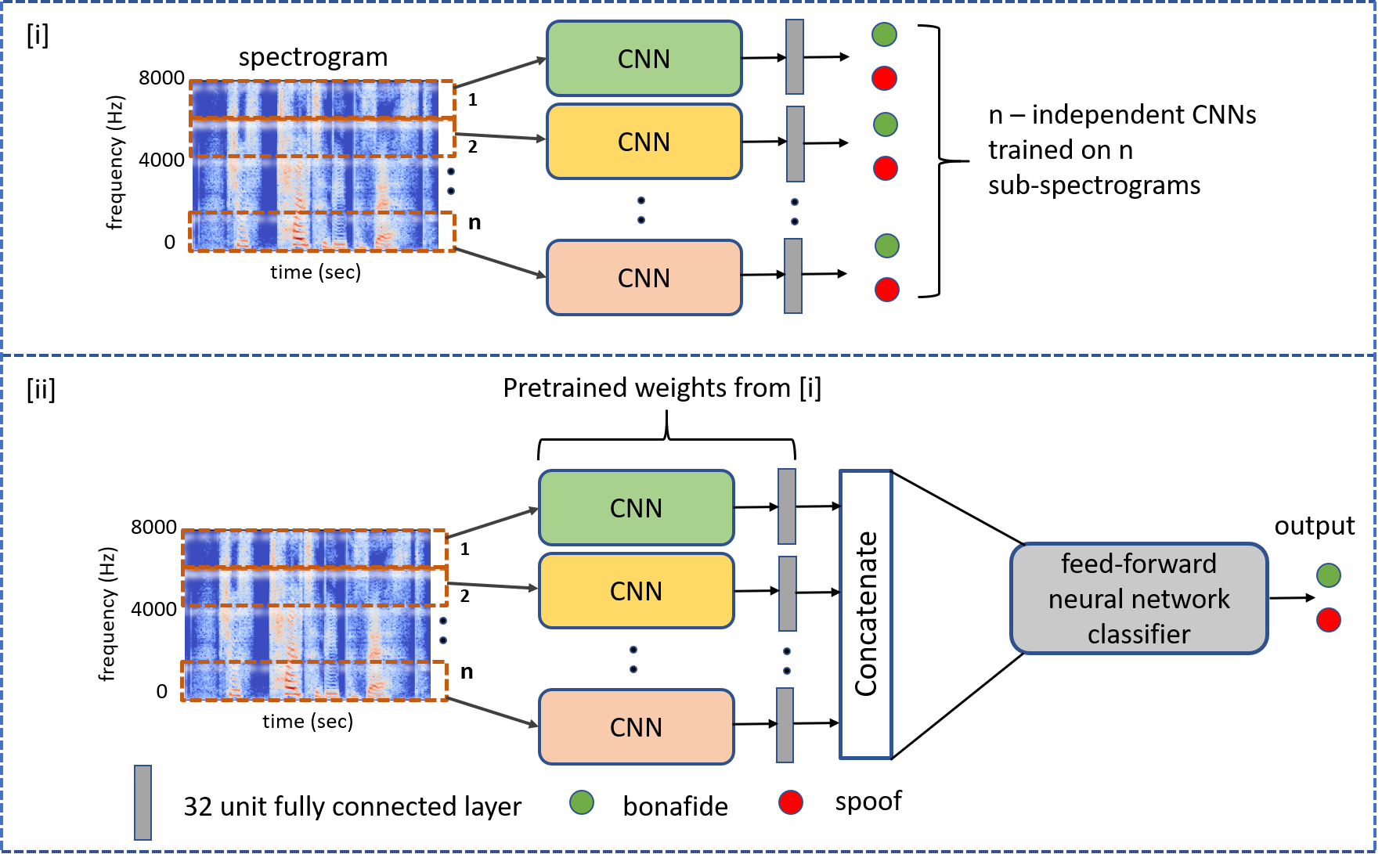}} 
	
	
	\caption{Proposed subband CNN modeling framework.} 
	\label{fig:subband-architecture}
\end{figure}

We focus on \textbf{feature extraction} for audio spoofing attack detection. There is a vast body of prior research on developing and enhancing different low-level feature extractors (most relevant work is reviewed in Section \ref{sec:related-work}), some of them obtaining very low spoof-bonafide detection error rates (even 0\%) on specific datasets. Many of these techniques leverage domain knowledge, whether speech science (speech production or perception), signal processing theory, or both. A potential benefit of such rationale is transparency and interpretability. At the same time, feature extractors crafted with the aid of domain knowledge might be too simplistic. As illustrated in Figure \ref{fig:subband-architecture} (b), we aim at hitting a suitable balance between hand-crafted and data-driven feature extraction: we use \textbf{spectrograms} (a meaningful representation of audio), processed in disjoint \textbf{subbands} (to divide-and-conquer high-dimensional spectrogram modeling across several, frequency-localized models, each handling a lower-dimensional feature space), each modeled with a \textbf{convolutional neural network} to learn band-specific features. The subband-specific features are concatenated to form feature vectors that are then classified with a feedforward neural network.

The general idea of processing a power spectrogram in subbands, as such, is not new in the speech field. Mel-frequency cepstral coefficients (MFCCs) are extracted using a filterbank consisting of frequency-localized subbands and subband-based modeling of speaker traits dates at least two decades back \cite{Besacier2000-subband}. In conventional, or \emph{fullband} models, one trains a single model (with a large number of parameters) using a descriptor of the fullband spectrum. In \emph{subband} based models, the rationale is to instead divide-and-conquer the task across independently modeled subbands which are later recombined using feature or score fusion techniques. The potential benefits include the possibility to side-step the `curse of dimensionality' by using a set of models trained on lower-dimensional inputs, to achieve robustness to \emph{frequency-selective noise}, and the possibility to analyze the importance (contribution) of each subband to the classification results. To this end, we summarise the main contributions of our work below.

\begin{itemize}
   \item We perform a systematic study on different subbands and their contribution in replay spoofing detection using convolutional neural networks (CNNs). 
   
    
   \item We propose a CNN-based subband modelling framework (Figure \ref{fig:subband-architecture} (b)) that offers substantial improvement over traditional fullband modelling (Section \ref{sec:results_discussion}).
   
   
   \item We study the effect of late fusion of subband CNNs using linear and weighted linear sum fusion approaches.

   \item Finally, we study the generalisability of all our replay spoofing countermeasures on the ASVspoof 2019 ``real'' PA test dataset. 
   
\end{itemize}

\section{Relation to prior work}\label{sec:related-work}

Since the release of publicly available benchmark datasets \cite{wu_IS2015,tomiSummaryPaper,asvspoof2019overview}, there is a substantial body of prior research on various types of spoofing countermeasures; refer to \cite{sahid_PAD_book} for a recent survey. The front-end processing part of spoofing countermeasures relies heavily on hand-crafted signal processing techniques. Besides the now-standard \emph{Constant Q cepstral coefficients} (CQCCs) \cite{hector_cqcc} (and extensions, e.g. \cite{rohan_eCQCC}), there are methods leveraging from modulation domain processing (e.g. \cite{tharshini_IS2018,patel_IS2017,madhu_IS2018}), adaptive filterbanks \cite{buddhi2019IS}, restricted Boltzmann machines \cite{hardik_IS2018}, and envelopes of subband signals \cite{buddhi_IS2018} to name a few. In this work, we focus on modeling spectral subbands obtained through a standard windowed discrete Fourier transform. 

There are some research works investigating the impact of subbands on model performance in different tasks. For instance, the authors in \cite{Besacier2000-subband} used a subband approach to extract relevant features, each modeled using Gaussians. Authors in \cite{subband_asv} investigated dependencies of different frequency bands and speaker characteristics in a speech signal for speaker verification applications. Recently, authors in \cite{emman_subnet} demonstrated improved performance using different subbands for building acoustic scene classification models. They use a \emph{sub-spectrogram}, obtained by cropping a mel-spectrogram at different bands, to train a convolutional neural network (CNN) for learning band-specific features. This is closely related to our work. One notable difference, however, is that we do not use overlapping bands. Furthermore, this approach has never been applied for spoofing detection tasks. 

In the context of spoofing detection, the most relevant studies include \cite{kavya_subband,Witkowski2017_subband,garg_subband_analysis,Nagarsheth2017_highfrequency,lin_subband_replay_attack,Soni_subband_spoofing}. The authors of \cite{kavya_subband} investigated different subbands to find the most informative bands useful for spoofing detection tasks. Using the Kullback-Leibler divergence (at model-level) and classification-level analysis, they identified 0-1 kHz, 2.5-5.5 kHz and 7-8 kHz as the most informative subbands on their dataset (SAS corpus). Features were then extracted from these bands to train a classifier (\emph{Gaussian mixture model --- universal background model}), demonstrating improved performance over traditional fullband models. The authors of \cite{Witkowski2017_subband} investigated the importance of different subbands for spoofing detection on the ASVspoof 2017 dataset. They extracted five different types of features from these subbands to train a GMM back-end classifier. They found the high frequency range of 6-8 kHz to be the most informative. The authors of \cite{garg_subband_analysis} similarly performed subband analysis using CQCC and MFCC features for replay spoofing detection on the ASVspoof 2017 dataset, with similar findings to those in \cite{Witkowski2017_subband}. The high frequency band (6-8 kHz) provides particularly relevant information for replay attack detection on this dataset. The authors in \cite{Nagarsheth2017_highfrequency} proposed \emph{high-frequency cepstral coefficient} (HFCC) features extracted from a high-frequency spectrum (above 3.5 kHz). They combined HFCCs with CQCCs and trained a deep neural network as a feature extractor. A support vector machine classifier was trained on deep features, outperforming the baseline GMM models on the ASVspoof 2017 dataset. The authors of \cite{Soni_subband_spoofing} trained a subband autoencoder (SBAE) for feature extraction by restricting connections between units in the input and the first hidden layer of the encoder. Doing so allows the model to learn band-specific features useful in spoofing detection, demonstrating substantial gain in detection performance on the ASVspoof 2015 dataset. Another line of study that investigates subbands for spoofing detection is \cite{lin_subband_replay_attack}. Using the subbands that provide discriminative information, they design new filters for feature extraction. Experimental results on the ASVspoof 2017 v1.0 dataset indicate that the $0$-$1$ and $7$-$8$ kHz subbands offer the most discriminative information.

To sum up, previous studies indicate that certain frequency subbands are potentially more informative to the detection of spoofing attacks, even though no standardized approach how that unevenly distributed information across the frequency axis should be utilized. Our current work is different from prior works  \cite{kavya_subband,Witkowski2017_subband,garg_subband_analysis,Nagarsheth2017_highfrequency,lin_subband_replay_attack} because most of them aim at hand-crafting or learning features \cite{Soni_subband_spoofing} based on the relevance of specific subbands for spoofing detection. To the best of our knowledge, there is no work in spoofing detection aiming to learn band-specific features by discriminatively training CNNs on a spectrogram input. This is one of the objectives of this paper. 

\section{Proposed methodology}\label{sec:proposed_method}
\emph{Convolutional neural network} (CNN) based countermeasure models trained using spectrograms have shown state-of-the-art performance in spoofing detection tasks in the ASVspoof 2017 challenge. They are usually trained using the fullband spectrum of the input signal and use a fixed-duration input representation \cite{galina_IS2017}. This \textit{conventional approach} of building CNN-based countermeasures is illustrated in Figure \ref{fig:subband-architecture} (a). As the CNN is trained discriminatively, it is forced to learn discriminative features using the entire frequency spectrum of the input signal, using a \emph{single} worker to extract usable information across all the frequency subbands for spoofing attack detection. 

But as the prior studies (Section~\ref{sec:related-work}) suggest, not all the subbands are necessarily equally informative. From a modeling perspective, the raw spectrogram patch (extracted by stacking multiple frames using all the frequency bands) is a high-dimensional vector, with strong correlations between any neighboring time (frame) or frequency (DFT bin) indices. As such low-level redundancy is common to both human and spoofed samples, it does not necessarily help in the discrimination (classification) task itself; instead, the model will have to learn both data compression (suppressing statistical redundancy to a useful intermediate representation) \emph{and} classification tasks. This may also result in additional computational time during convolution operations.

Therefore, rather than having a single CNN that merges information across different frequency bands, we propose to incorporate a \emph{bank} of $n$ different CNNs, each operating on non-overlapping $n$ frequency subbands. Our proposed methodology is illustrated in Figure~\ref{fig:subband-architecture} (b). Each of the \emph{subband CNNs} now has to model a much lower-dimensional subspace, producing a less redundant and more relevant representation of its respective subband. Note that the subband representations are afterwards re-combined through concatenation. This new representation now contains again information across the full frequency band, allowing any subsequent model to exploit possibly useful band-level correlations. In our case, we use a simple feedforward neural network (FFNN) for the final classification. Natural questions that arise now are \emph{how to perform the frequency-domain split} and \emph{how to choose $n$.} We address three different forms of splits, $n=2$, $n=4$ and $n=8$, and go for the easiest choice of uniform frequency division.

This choice is motivated from \cite{lin_subband_replay_attack}. They divide the original spectrogram into $n$ uniform subbands with bandwidth $1$ kHz corresponding to $n=8$ splits and $0.5$ kHz bandwidth for $n=16$ splits. They remove one subband at a time and hand-craft features from the remaining subbands. GMMs are then trained on these features for spoofing detection and the performance is evaluated in terms of EER on the ASVspoof 2017 evaluation sets. As our dataset consists of 16 kHz audio (Nyquist range 8 kHz), our three choices correspond to subbands of bandwidths 4 kHz ($n=2$), 2 kHz ($n=4$) and 1 kHz ($n=8$). It should be noted that the default case $n = 1$ corresponds to the baseline CNN (Figure~\ref{fig:subband-architecture} a), i.e the model trained on the fullband spectrogram. From hereon we use ``CNN'' to refer to the baseline CNN. And, we use ``sub-CNN" to refer to models trained on the subband spectrograms.

We operate on power spectrograms instead of other alternative time-frequency representations, following findings in \cite{galina_IS2017}. All our sub-CNNs use the architecture described in \cite{bhusanSLT2018}, which is an adapted version of the best performing model \cite{galina_IS2017} in the ASVspoof 2017 challenge. It consists of $9$ convolutional layers, $5$ max-pooling layers and $2$ fully connected (FC) layers. The key difference while training such sub-CNNs is in terms of the input they receive. The bandwidth of the input sub-spectrogram varies depending upon different values of $n$ (number of splits). Section~\ref{sec:setup} provides more details regarding input representations, training and testing of these models. 

The proposed joint sub-CNN model of Figure~\ref{fig:subband-architecture} (b) second row uses the same architecture as in sub-CNNs (first row of the same figure) with the following updates: (1) there is no output layer now, and (2) a concatenation layer is added that merges the fully-connected (FC) layer output from $n$ sub-CNN models producing a $32 \times n$ dimensional vector. It should be noted that the choice of $32$ units in the FC layer comes from the baseline CNN of Figure~\ref{fig:subband-architecture} (a). Furthermore, this architecture with $32$ FC units has shown promising results as described in \cite{bhusanSLT2018}. Next, the concatenated vector is fed to a feedforward neural network (FFNN) for class discrimination. The FFNN consists of two fully-connected layers with $256$ and $128$ units. This is followed by a single unit output layer with sigmoid non linearity for class discrimination. We apply batch normalisation before applying ReLU non-linearity to these layers. The architecture of the FFNN is optimised through model validation (on the development set). Training and optimisation of our proposed framework is done in two steps:

\begin{itemize}
	\item First, the input spectrogram is split into $n$ non-overlapping sub-spectrograms and $n$ sub-CNNs are trained independently on them. The training dataset is used for model training and the development dataset is used for model validation. This step is depicted in the top row of Figure~\ref{fig:subband-architecture}(b).
	
	\item Second, the pretrained sub-CNNs (excluding the last layer) are used to initialise the weights of the sub-CNN modules of our joint sub-CNN framework shown in the bottom row of Figure~\ref{fig:subband-architecture} (b). The weights of the FFNN layers are initialised randomly using xavier initialization \cite{glorot}. The biases are initialised to zero. Given an input spectrogram, the framework first splits it into $n$ non-overlapping sub-spectrograms which are processed by $n$ sub-CNNs and the whole network parameters are jointly updated during backpropagation. This step can be interpreted as fine-tuning of the subband CNNs and the classifier back-end jointly for best performance. As in the earlier step, model parameters are trained on the training dataset and the development set is used for model validation. 
\end{itemize}

Our proposed work is different from prior works  \cite{kavya_subband,Witkowski2017_subband,garg_subband_analysis,Nagarsheth2017_highfrequency,lin_subband_replay_attack} because most of them aim at hand-crafting or learning features \cite{Soni_subband_spoofing} based on the relevance of specific subbands for spoofing detection. To the best of our knowledge, there is no work in spoofing detection aiming to learn band-specific features by discriminatively training CNNs on a spectrogram input.

\section{Experimental setup}\label{sec:setup}

Here we describe our experimental setup including the details of dataset used, model architecture, training and scoring procedures and also the evaluation measures.

\subsection{Dataset}\label{ssec:dataset}
We use two publicly available spoofing datasets, \textbf{ASVspoof 2017 v2.0} \cite{hectorAsvspoof2.0} and \textbf{ASVspoof 2019 physical access} (PA), \cite{asvspoof2019overview} for model training and testing. In addition, we also include results on the recently released \textbf{ASVspoof2019 real PA} dataset\footnote{\url{https://www.asvspoof.org/database}} for the challenging case of cross-database performance evaluation. All the datasets are representative of \emph{replay attacks} and are complementary to each other.

The ASVspoof 2017 v2.0 dataset consists of bonafide and spoof class audio recordings grouped into three subsets: training, development and evaluation \cite{tomiSummaryPaper,hectorAsvspoof2.0}. It contains male speakers only: $10$, $8$ and $24$ in the training, development and evaluation sets, respectively. The training set contains a balanced set of audio files, $1507$ for both bonafide and spoof classes. The development set has $760$ bonafide and $950$ spoof files. The evaluation set has $1298$ bonafide and $12008$ spoof audio files \cite{hectorAsvspoof2.0}.

The ASVspoof 2019 PA dataset, in turn, consists of $8$ male and $12$ female speakers in the training and development subsets. Both the training and development sets have $5,400$ bonafide utterances, while $48,600$ and $24,300$ spoofed utterances are included in the training and development sets, respectively. The evaluation set has $135,000$ test utterances \cite{asvspoof2019overview,asvspoof2019_evaluationplan}. Following our prior findings \cite{bhusan2019challenge} on the ASVspoof 2019 dataset, we adopt a custom, but publicly available protocol\footnote{https://github.com/BhusanChettri/ASVspoof2019}. 

While the ASVspoof 2019 PA dataset was created using \emph{simulated} replay attacks, the ASVspoof2019 real PA dataset consists of audio recordings developed under real replay conditions. We use this last data to evaluate our models (trained using ASVspoof 2017 and 2019 PA data) to gauge cross-dataset performance. The real PA dataset consists of $2700$ audio files with $540$ bonafide and $2160$ spoof recordings \cite{asvspoof2019overview}. 

\subsection{Input representation and preprocessing} \label{input_and_preprocessing}
The input to the network is a mean-variance normalized log power spectrogram of $3$ seconds. This normalisation, motivated from \cite{galina_IS2017}, is performed at the utterance-level to standardize the features (zero mean and unit variance for all frequency bins) within a given recording. We use a 512-point \emph{fast Fourier transform} (FFT), and a 32 ms window with a hop of 10 ms. Therefore, the original input spectrogram has a shape of $300\times257$, where 300 is the number of frames and 257 the number of FFT bins. To obtain a consistent input representation we replicate the audio samples (in the time domain) if the duration is smaller, or truncate the samples to 3 seconds duration. When $n=1$, the input shape to the baseline CNN remains the same ($300\times257$) however it varies for sub-CNNs depending upon different splits we use. For example, when $n$ = 2, the input shape becomes $300\times128$ and $300\times129$. We always include the leftover bin to the last split. Similarly, when $n$ = 4 we have four sub-spectrograms where the first three will have a shape of $300\times64$, and a shape of $300\times65$ for the last split. Likewise, for $n$ = 8 we have seven sub-spectrograms of shape $300\times32$, and a shape of $300\times33$ for the last split.  

Following our prior findings \cite{bhusan2019challenge} on the ASVspoof 2019 PA dataset, we remove zero-valued samples from the start and end of every audio recording in the dataset. Likewise, on the ASVspoof 2017 v2.0 dataset, we remove leading and trailing silence/nonspeech samples following our findings in \cite{bhusanSLT2018}. For this, we use our publicly released speech endpoint annotations \cite{bhusan_chettri_2020_3601188}. Applying such preprocessing helps the model to avoid exploiting cues that are actually not relevant to the problem, rather this forces the models to now learn relevant factors in replay spoofing detection.

\subsection{Model training and testing}\label{ssec:training_testing}

We train the network to optimize the binary cross entropy loss between a bonafide and a spoof class. We use a batch size of $32$ and learning rate of $1e^{-4}$. We use the \emph{ADAM} \cite{adam} optimizer with default parameters. We apply a dropout of 50\% to the input of the fully connected layers. If the validation loss does not improve for $5$ epochs we stop the training process to avoid overfitting. We train models for a maximum of $100$ training epochs. Using this approach we train $5$ models with random initialisation. We choose the model showing the best performance on the development set and use it to test the performance on the evaluation set. At test time, for each audio spectrogram  we use the model output --- the bonafide-class posterior probability --- as our detection score. The approach described above is the same for all our models. 


\subsection{Performance evaluation}

We assess the performance of different countermeasure models using two different metrics. The first one gauges the ability of the countermeasure to discriminate bonafide and spoofed utterances from each other. It is measured through the \emph{equal error rate} (EER), which was the primary evaluation metric of the ASVspoof 2017 challenge, and a secondary metric of the ASVspoof 2019 challenge. EER is the error rate at an operating point where the false acceptance (\emph{false alarm}) and false rejection (\emph{miss}) rates are equal. A reference value of 50\% indicates the chance level. Our second metric is  \emph{minimum normalized tandem detection cost function} (t-DCF) \cite{tomi_tDCF} metric that evaluates both countermeasure and ASV system as a whole. We use the evaluation scripts and the ASV scores (x-vector recognizer) released by the organisers. A reference value 1.00 of (normalized) t-DCF indicates an uninformative countermeasure.


\section{Experiments and results}\label{sec:results_discussion}
We now describe different experimental studies along with the results in this section. \\

\begin{table}
	\caption{Performance of the baselines.}
	\centering
	\begin{tabular}{cccccc}
		\hline
		\multirow{2}{*}{Baseline}& \multirow{1}{*}{Subbands}&\multicolumn{2}{c}{ASVspoof 2017} &\multicolumn{2}{c}{ASVspoof 2019}  \\
		& (kHz) &t-DCF & EER\% & t-DCF & EER\%  \\
		\hline
		CNN & $0$-$8$ &$0.3873$ &$13.02$ &$0.2019$ &$7.00$ \\
		GMM & $0$-$8$ &$0.5054$ &$18.33$ &$0.9928$ &$59.12$ \\
		\hline
	\end{tabular}
	\label{baseline_results}
\end{table}

\noindent
\textbf{Baselines}. To assess the performance of our proposed framework we train a baseline CNN model (Figure \ref{fig:subband-architecture}(a)) using a fullband spectrogram. For completeness, we also train and test a CQCC-based GMM model so as to compare it with our proposed framework. It should be noted that due to the preprocessing applied on the ASVspoof 2017 and 2019 PA datasets during training and testing, our baselines are different from the official ones \cite{hectorAsvspoof2.0,asvspoof2019overview}; therefore, the numbers reported here should not be directly compared with the results of ASVspoof challenges. Table \ref{baseline_results} summarises the results.

On the ASVspoof 2017 v2.0 we use the same parameterisation as in \cite{hectorAsvspoof2.0} to train the GMM model. On the evaluation set, the GMM model reaches EER = $18.33$\% and t-DCF = $0.5054$. On the ASVspoof 2019 PA\footnote{We used the pretrained GMM model from our prior work \cite{bhusan2019challenge} to test on the evaluation set.} evaluation set EER = $59.12$\% and t-DCF = $0.9928$. The results suggest that GMM models become less confident in making classification decisions when silence cues are removed during training and testing. Refer to \cite{bhusan2019challenge} for more details. On both datasets, the CNN baseline outperforms the GMM on both metrics. This demonstrates its effectiveness on learning relevant features useful for discrimination despite the preprocessing applied on the audio signals in contrast to hand-crafted CQCC features used in the GMM.


\begin{table}
	\caption{Performance of sub-CNNs M$_1$, M$_2$ and joint model J$_1$. Bold indicates the best performance.}
	\centering
	\begin{tabular}{cccccc}
		\hline
		\multirow{2}{*}{Model}& \multirow{1}{*}{Subbands}&\multicolumn{2}{c}{ASVspoof 2017} &\multicolumn{2}{c}{ASVspoof 2019}  \\
		& (kHz) &t-DCF & EER\% & t-DCF & EER\%  \\
		\hline
		M$_1$ & $0$-$4$ &$0.7045$ &$33.03$ &$0.1925$ &$6.97$ \\
		M$_2$ & $4$-$8$ &$0.4800$ &$18.95$ &$0.5201$ &$19.69$  \\
		J$_1$ &- &$\textbf{0.2893}$ &$\textbf{10.63}$ &$\textbf{0.1864}$ &$\textbf{6.44}$ \\
		\hline
	\end{tabular}
	\label{experiments1_results_part1}
\end{table}

\begin{table}
	\caption{Performance of sub-CNNs M$_3$ through M$_6$ and joint model J$_2$. Bold indicates the best performance.}
	\centering
	
	\begin{tabular}{cccccc}
		\hline
		\multirow{2}{*}{Model}& \multirow{1}{*}{Subbands}&\multicolumn{2}{c}{ASVspoof 2017} &\multicolumn{2}{c}{ASVspoof 2019}  \\
		& (kHz) &t-DCF & EER\% & t-DCF & EER\%  \\
		\hline
		M$_3$ & $0$-$2$ &$0.6172$ &$31.20$ &$0.2265$ &$7.97$ \\
		M$_4$ & $2$-$4$ &$0.9773$ &$41.58$ &$0.4568$ &$17.72$\\
		M$_5$ & $4$-$6$ &$0.9995$ &$50.70$ &$0.5903$ &$23.16$ \\
		M$_6$ & $6$-$8$ &$0.5032$ &$23.50$ &$0.6019$ &$23.59$ \\
		
		J$_2$ &- &$\textbf{0.3343}$ &$\textbf{11.78}$ &$\textbf{0.1977}$ &$\textbf{6.99}$ \\
		\hline
	\end{tabular}
	\label{experiments1_results_part2}
\end{table}

\subsection{Experiment 1: subband modeling}\label{ssec:experiments1}

We design four experimental setups for different values of $n$ using our proposed methodology. We use $n=2$ in our first setup. Using the architecture and training methodology described earlier we train two independent sub-CNNs M$_1$ and M$_2$. M$_1$ operates on the first $4$ kHz and M$_2$ on the last $4$ kHz subband spectrograms. Next, we use them (except the last output layer) to initialise the respective sub-CNN module weights of our joint sub-CNN model framework as shown in the bottom row of Figure \ref{fig:subband-architecture}(b). We call this joint model J$_1$. Our second setup uses $n=4$. Therefore, we train four independent sub-CNNs M$_3$ through M$_6$ on $2$ kHz subband spectrograms. We then use these pretrained models to initialise the weights of sub-CNN modules of our joint model J$_2$. Our final setup uses $n=8$. We now train eight independent sub-CNNs M$_7$ through M$_{14}$ operating on $1$ kHz subband spectrograms. We then use them (except the last layer weights) to initialise our joint model J$_3$ shown in Figure \ref{fig:subband-architecture}(b) bottom row. 

Finally, motivated from the results of sub-CNNs M$_7$ and M$_{14}$ on the ASVspoof 2017 dataset (shown in Table~\ref{experiments1_results_part3}) we design a joint model J$_4$ operating on the first and the last $1$ kHz subband spectrograms. This setup is different from the previous setups as more than half of the information is being discarded here. Overall J$_4$ utilises only $2$ kHz of information (the first and last $1$ kHz bands). As in the earlier setups, the pretrained weights (here M$_7$ and M$_{14}$ models) are used to initialise the weights of sub-CNN modules of our joint model J$_4$. It should be noted that the entire model parameters are jointly optimised while training J$_1$, J$_2$, J$_3$ and J$_4$. We perform these experiments on both the ASVspoof 2017 v2.0 and 2019 PA datasets.

Table \ref{experiments1_results_part1} summarises the performance of our individual sub-CNNs M$_1$, M$_2$ and the joint model J$_1$. On the ASVspoof 2017 dataset, the higher frequency bands ($4-8$ kHz) seem to carry more discriminative information than the lower bands ($0-4$ kHz). However, we observe the opposite pattern on the ASVspoof 2019 dataset where M$_1$ trained on $0-4$ kHz shows better results than M$_2$. This might be due to differences in dataset design (real vs. simulated replay) and compilation (different speakers and audio qualities of the source corpora). Nonetheless, on both the datasets our proposed joint model J$_1$ outperforms M$_1$, M$_2$ and the baselines by a large margin on both performance metrics. This demonstrates the effectiveness of our proposed approach for spoofing attack detection.

Next we discuss the performance of sub-CNNs M$_3$ through M$_6$ and the joint model J$_2$. Table \ref{experiments1_results_part2} summarises this. Overall, the individual sub-CNNs now show poor performance, which is expected as each of them now receives only half of the information as the previous case ($n=2$). Nonetheless, the proposed model J$_2$ that merges information from all the bands again outperforms the fullband baselines (CNN and GMM) and sub-CNNs M$_3$ through M$_6$.

Table \ref{experiments1_results_part3} summarises the performance of our individual sub-CNNs M$_7$ through M$_{14}$ and joint models J$_3$ and J$_4$. These results provide deeper insights in understanding the influence of different subbands for spoofing detection. As in the previous two setups ($n=2$ and $n=4$), our joint model J$_3$ shows better results than training individual sub-CNNs indicating that joint training offers some form of complementary information across different frequency bands. We also find that on the ASVspoof 2017 dataset, M$_{14}$ trained on the \emph{last} $7-8$ kHz band outperforms all other sub-CNNs by a large margin. This is followed by M$_{7}$ (operating on the $0-1$ kHz band) that also performs substantially better than the remaining sub-CNNs. Inspired by this finding, we further train another joint model J$_4$ trained \emph{only} with the lowest and highest sub-CNNs. Interestingly, this highly reduced model yields the best performance on the ASVspoof 2017 dataset, matching with the findings reported on the version 1.0 dataset by \cite{lin_subband_replay_attack}. However, on the ASVspoof 2019 dataset, the first $1$ kHz subband appears to carry most relevant information as opposed to other subbands. Furthermore, the results found for the ASVspoof 2017 and 2019 dataset do not match completely with the main differences being in dataset design and collection --- the two datasets ASVspoof 2017 and 2019 PA are designed and collected differently. 

\begin{table}
	\caption{Performance of sub-CNNs M$_7$ through M$_{14}$ and joint models J$_3$ and J$_4$. Bold indicates the best performance.}
	\centering
	
	\begin{tabular}{cccccc}
		\hline
		\multirow{2}{*}{Model}& \multirow{1}{*}{Subbands}&\multicolumn{2}{c}{ASVspoof 2017} &\multicolumn{2}{c}{ASVspoof 2019}  \\
		& (kHz) &t-DCF & EER\% & t-DCF & EER\%  \\
		\hline
		M$_7$ & $0$-$1$ &$0.6216$ &$31.59$ &$0.2354$ &$8.48$ \\
		M$_8$ & $1$-$2$ &$0.9977$ &$48.92$ &$0.6557$ &$25.15$\\
		M$_9$ & $2$-$3$ &$0.9971$ &$50.61$ &$0.7327$ &$28.60$ \\
		M$_{10}$ & $3$-$4$  &$0.9969$ &$49.85$ &$0.5390$ &$21.60$ \\
		M$_{11}$ & $4$-$5$ &$0.9992$ &$44.06$ &$0.6605$ &$25.69$ \\		
		M$_{12}$ & $5$-$6$ &$0.9914$ &$45.05$ &$0.7118$ &$28.34$ \\
		M$_{13}$ & $6$-$7$ &$0.9817$ &$42.23$ &$0.7123$ &$28.17$ \\	    
		M$_{14}$ & $7$-$8$ &$0.4747$ &$18.10$ &$0.7231$ &$28.01$ \\
		
		J$_3$ &- &$\textbf{0.2734}$ &$11.02$ &$\textbf{0.1975}$ &$\textbf{7.34}$ \\
		\hline
		J$_4$ &- &$0.2771$ &$\textbf{10.40}$ &$0.2238$ &$8.30$ \\
		\hline
		
	\end{tabular}
	\label{experiments1_results_part3}
\end{table}

\subsection{Experiment 2: score fusion}\label{ssec:experiments2}

\begin{table}
	\caption{Score-level fusion results. LS: linear sum of scores. WLS: weighted sum of scores.}
	\centering
	
	\begin{tabular}{cccccc}
		\hline
		\multirow{2}{*}{Model}& \multirow{1}{*}{Fusion}&\multicolumn{2}{c}{ASVspoof 2017} &\multicolumn{2}{c}{ASVspoof 2019}  \\
		& (type) &t-DCF & EER\% & t-DCF & EER\%  \\
		\hline		
		F$_1$ & LS &$0.3208$ &$11.86$ &$0.2208$ &$7.50$ \\
		F$_2$ & WLS  &$0.3189$ &$11.72$ &$0.2034$ &$\textbf{6.78}$  \\
		F$_3$ & LS &$0.6151$ &$24.25$ &$0.2197$ &$8.00$ \\
		F$_4$ & WLS  &$\textbf{0.3079}$ &$\textbf{11.55}$ &$0.1898$ &$6.95$ \\
		F$_5$ & LS &$0.4932$ &$18.25$ &$0.2626$ &$9.78$ \\
		F$_6$ & WLS  &$0.3548$ &$12.33$ &$\textbf{0.1848}$ &$6.99$   \\
		
		\hline
	\end{tabular}
	\label{fusion_results}
\end{table}

\begin{table}
	\caption{Cross-dataset performance evaluation on the 2019 PA real testset. D1: ASVspoof 2017 v2.0, D2: ASVspoof 2019 PA.} 
	
	\centering
	\begin{tabular}{cccccc}
		\hline
		\multirow{2}{*}{Model}& \multirow{1}{*}{Subbands}&\multicolumn{2}{c}{Trained on D1} &\multicolumn{2}{c}{Trained on D2}  \\
		& (kHz) &t-DCF & EER\% & t-DCF & EER\%  \\
		\hline
		CNN & $0$-$8$ &$0.9986$ &$41.29$ &$0.6374$ &$34.25$ \\
		
		M$_1$ & $0$-$4$ &$0.8205$ &$44.81$ &$0.6800$ &$33.12$ \\
		M$_2$ & $4$-$8$ &$0.9687$ &$42.59$ &$0.7820$ &$39.44$ \\
		J$_1$ &- &$0.9939$ &$44.07$ &$0.6741$ &$35.92$ \\
		\hline		
		
		M$_3$ & $0$-$2$ &$0.8984$ &$38.33$ &$0.6760$ &$33.49$ \\
		M$_4$ & $2$-$4$ &$\textbf{0.8025}$ &$46.08$ &$0.7311$ &$34.62$ \\
		M$_5$ & $4$-$6$ &$0.9995$ &$59.44$ &$0.8033$ &$35.74$ \\
		M$_6$ & $6$-$8$ &$0.9976$ &$51.85$ &$0.6877$ &$28.37$ \\
		J$_2$ &- &$0.9944$ &$40.55$ &$0.6544$ &$30.18$ \\
		\hline
		
		M$_7$ & $0$-$1$ &$0.8508$ &$\textbf{34.81}$ &$0.6678$ &$34.95$ \\
		M$_8$ & $1$-$2$ &$0.9875$ &$47.63$ &$1.0$ &$51.29$ \\
		M$_9$ & $2$-$3$ &$0.9545$ &$51.48$ &$0.8014$ &$32.22$ \\
		M$_{10}$ & $3$-$4$ &$0.9634$ &$42.22$ &$0.8553$ &$35.41$ \\
		M$_{11}$ & $4$-$5$ &$0.9003$ &$43.33$ &$0.7524$ &$32.80$ \\	
		M$_{12}$ & $5$-$6$ &$0.9788$ &$47.03$ &$0.8322$ &$38.65$ \\
		M$_{13}$ & $6$-$7$ &$0.9828$ &$48.49$ &$0.7038$ &$28.70$ \\    
		M$_{14}$ & $7$-$8$ &$0.9981$ &$42.38$ &$\textbf{0.6483}$ &$\textbf{27.22}$ \\		
		J$_3$ &- &$0.9208$ &$37.61$  &$0.6439$ &$30.97$  \\
		
		J$_4$ &- &$0.9081$ &$36.62$  &$0.7541$ &$36.71$  \\
		\hline
		
	\end{tabular}
	\label{experiments1_results_part4}
\end{table}

We perform fusion experiments to understand if combining information through score-level fusion helps improve detection performance. In this setting we use the scores from each of the pretrained sub-CNNs and combine their scores using two simple approaches: linear sum of scores (LS), and linear weighted sum of scores (WLS). Let $S_1, S_2, S_3, \text{...}, S_n$ represent scores from $n$ sub-CNNs for a test utterance $X$. The fused score using the two approaches is obtained as: $\text{linear sum} = S_1 + S_2 + S_3 + .. + S_n$ and $\text{weighted sum} = w_1*S_1 + w_2*S_2 + w_3*S_3 + .. + w_n*S_n$ where $w_1, w_2, w_3, \text{...}, w_n$ are weights corresponding to each sub-CNN score learned using logistic regression (LR). We use the Bosaris toolkit \cite{bosaris2013brummer} for the LR implementation. We perform six different fusion experiments: F$_1$ - F$_6$. F$_1$ uses linear sum fusion on the M$_1$ and M$_2$ model scores, F$_2$ on the other hand uses weighted linear fusion on them. F$_3$ uses linear sum fusion on the scores of the models M$_3$ through M$_6$, F$_4$ on the other hand uses weighted linear fusion on them. F$_5$ uses linear sum fusion on the scores of the models M$_7$ through M$_{14}$, F$_6$ on the contrary applies weighted linear fusion on them. 

Table \ref{fusion_results} summarises the results. In general, WLS fusion shows better performance than LS fusion, outperforming the baselines. However they show poor (or similar) performance compared to our joint subband modeling framework. For example, on the ASVspoof 2017 evaluation set the joint model J$_4$ (Table \ref{experiments1_results_part3}) shows better results in comparison to all our fusion models. Similarly, on the ASVspoof 2019 PA evaluation set the joint model J$_1$ (Table \ref{experiments1_results_part1}) shows better EER (and slightly worse t-DCF). As expected, combining information through score-fusion techniques offers gains in detection performance. However, our proposed joint subband modelling framework shows better results over score-fusion approaches. This further confirms that the complementary information provided by individual sub-CNNs helps improve overall detection performance. 

\subsection{Experiment 3 - cross-database evaluation}\label{ssec:experiments3}

The goal of our final experiment is to understand the generalisability of our subband specific models (M$_1$ through M$_{11}$) and joint subband models (J$_1$, J$_2$, J$_3$ and J$_4$) in \emph{cross-database} evaluation. We take all our pretrained models trained on the ASVspoof 2017 and ASVspoof 2019 PA dataset and evaluate their performance on the ASVspoof 2019 real PA test set. Table \ref{experiments1_results_part4} shows the results.

While our models show good performance on the respective evaluation sets (see Tables \ref{experiments1_results_part1}, \ref{experiments1_results_part2}, \ref{experiments1_results_part3}), they fail to generalise well on the unseen real replay attack conditions of the ASVspoof 2019 real PA test set. On both the t-DCF and EER we observe high error rates for our joint models (J$_1$, J$_2$, J$_3$ and J$_4$). This also indicates overfitting on the respective dataset and lack of generalisability in unseen attack conditions. One possible interpretation to this observation may be attributed to dataset design and collection for the ASVspoof 2017 and 2019 PA datasets. It is worth noting that the ASVspoof 2019 PA dataset was developed through controlled simulation while the ASVspoof 2017 dataset was collected in real world recording and replay conditions. Due to the dataset issues \cite{bhusanSLT2018,bhusan2019challenge} identified on both ASVspoof 2017 and 2019 PA, the models trained on these datasets might not be able to capture real replay attack conditions and thus perform poorly on the real PA test set which has been designed carefully reflecting real replay attack conditions. This study further suggests that there is still need for a reliable replay training dataset that can be used to train models incorporating real world replay attack conditions.

\section{Discussion and conclusion}\label{sec:conclusion}

\begin{table}
	\caption{Summary of results showing the comparison of baselines with our proposed models. * results taken from Table~\ref{baseline_results}.}
	
	\centering
	\begin{tabular}{ccccc}
		\hline
		\multirow{2}{*}{Model}&\multicolumn{2}{c}{ASVspoof 2017} &\multicolumn{2}{c}{ASVspoof 2019}  \\
		&t-DCF & EER\% & t-DCF & EER\%  \\
		\hline		
		GMM*  &$0.5054$ &$18.33$ &$0.9928$  &$59.12$ \\
		CNN*  &$0.3873$ &$13.02$ &$0.2019$ &$7.0$ \\
		\hline
		J$_1$  &$0.2893$ &$10.63$ &$\textbf{0.1864}$ &$\textbf{6.44}$ \\
		J$_2$  &$0.3343$ &$11.78$ &$0.1977$ &$6.99$ \\
		J$_3$  &$\textbf{0.2734}$ &$11.02$ &$0.1975$ &$7.34$ \\
		J$_4$  &$0.2771$ &$\textbf{10.40}$ &$0.2238$ &$8.30$ \\
		\hline
	\end{tabular}
	\label{summary_of_results}
\end{table}

In this paper we performed a detailed analysis on the impact of different subbands and their importance on replay spoofing detection tasks on the benchmark datasets ASVspoof 2017 v2.0 and ASVspoof 2019 PA. Our proposed subband CNN model outperformed the traditional fullband CNN model, and also the CQCC-GMM baselines by a large margin, demonstrating the significance of our approach. We also investigated how combining information at the score level from each subband CNNs compared with our joint subband modelling framework. The final set of studies we performed here is on cross-database evaluation to investigate the generalisability of replay spoofing countermeasures on the ASVspoof 2019 real PA dataset.

Table \ref{summary_of_results} provides a summary of our main findings on both the ASVspoof 2017 and 2019 PA datasets. Performance improvement obtained using our proposed approach over baslines are very promising, encouraging further research on subband modelling for spoofing detection. Furthermore, it should be noted that our main objective in this paper is not to beat the best performing models published on the two datasets, but to validate our hypothesis about subband modelling using CNNs trained on spectrogram inputs. An important take-home message from our study is that while the first and the last 1 kHz frequency bands offer substantial improvement in detection performance, the same does not hold true for models trained on ASVspoof 2019. This suggests that the datasets available for training these models do not reflect real world replay conditions, suggesting a need for careful design of datasets for training replay spoofing countermeasures.

As future work, we aim to extend this analysis on text-to-speech and voice-converted spoofing attacks using the ASVspoof 2015 and ASVspoof 2019 logical access spoofing datasets. Use of advanced backend classifiers (current work uses a simple feed forward neural network) in our joint subband modelling framework will also be investigated. Furthermore, we are also interested in investigating the design and implementation of a unified countermeasure model that can be used for any spoofing attack conditions. 

\bibliographystyle{IEEEbib}
\bibliography{mybibliography}

\end{document}